\newcommand{\roughly}[1]%
    {{\mathrel{\raise.3ex\hbox{$#1$\kern-.75em\lower1ex\hbox{$\sim$}}}}}

\documentclass[12pt]{article}
\usepackage{paper2e}
\usepackage{latexsym}

\begin{document}
\begin{titlepage}

\preprint{IPMU13-0057}
\title{
 Emergence of time in power-counting renormalizable Riemannian theory of 
 gravity}
\author{Shinji Mukohyama}
\address{
 Kavli Institute for the Physics and Mathematics of the Universe\\
 Todai Institutes for Advanced Study, University of Tokyo\\
 5-1-5 Kashiwanoha, Kashiwa, Chiba 277-8583, Japan}

\begin{abstract}
 We suggest a new scenario of gravitation in which gravity at the
 fundamental level is described by a Riemannian (i.e. locally Euclidean)
 theory without the notion of time. The Lorentzian metric structure and
 the notion of time emerge as effective properties at long distances. On
 the other hand, at short distances, higher derivative terms compatible
 with the Riemannian diffeomorphism become important and the system is
 described by a power-counting renormalizable Riemannian theory. 
\end{abstract}

\end{titlepage}

\section{Introduction}
\label{sec:introduction}

Reconciliation of gravity and quantum theory is one of the most
outstanding problems in modern physics. Although general relativity (GR)
has been successful to reproduce and predict gravitational phenomena
over a vast range of scales, attempts to quantize GR lead to a number of
problems. In particular, if we naively applied the perturbative quantum
field theory approach to GR then we would end up with infinite number of
counter terms and would lose predictability at short distances such as
the Planck length. For this reason, various short-distance modifications
to GR have been proposed in the literature.

Inclusion of higher derivative terms is one of such modifications. 
Actually, if forth-order derivative terms in the action dominate the
system at short distances then one can render the theory of gravity
power-counting renormalizable. In the literature of forth-order gravity
theories, some evidences have been reported for
renormalizability~\cite{Stelle:1976gc}, for asymptotic freedom of
dimensionless couplings~\cite{Avramidi:1985ki,deBerredoPeixoto:2004if}
and for asymptotic safety of Newton's constant and the cosmological
constant~\cite{Codello:2006in}. These theories, however, are known to
have a serious problem: because of higher time derivatives present in
the action, it is not obvious how to maintain unitarity. There is an
attempt to remedy this problem by using the $1/p^2$-type propagator and 
including infinite number of interaction terms~\cite{Gomis:1995jp}, but
beta functions for this infinite set of couplings is not known at
present and it is expected that not all couplings are asymptotically
safe. The challenge is then how to reconcile asymptotically safe
couplings with unitarity. It is thus fair to say that the issue of
non-unitarity in higher derivative theories has not been
settled~\cite{review-AS}.

The purpose of this paper is to point out a possible alternative way to
get around the issue of non-unitarity while maintaining
renormalizability. This new possibility is based on two ideas: (i) in
Riemannian (i.e. locally Euclidean) theories with positive definite
metrics there is no dynamics and thus higher derivative terms do not
necessarily lead to a problem; (ii) the Lorentzian metric structure that
we usually consider as fundamental may be an effective property that
emerges only at long distances and only in some
regions~\cite{Mukohyama:2013ew}. We thus propose a scenario in which
gravity at short-distances is described by a power-counting
renormalizable Riemannian theory of gravity with a positive definite
metric and the Lorentzian metric structure emerges at long distances.

The rest of this paper is organized as follows. In Sec.~\ref{sec:theory}
we describe a power-counting renormalizable Riemannian theory of
gravity that potentially leads to emergence of time at long
distances in some regions. This theory contains a Riemannian metric with
positive definite signature and a clock field, i.e. a scalar field
playing the role of time by means of spontaneous symmetry breaking of
Riemannian diffeomorphism. In Sec.~\ref{sec:IRaction} we consider the
infrared (IR) limit of the theory and obtain the IR action. It is then
shown in Sec.~\ref{sec:LorentzianAction} that the IR action can be cast
into the form of an effective action for a Lorentzian theory so that the 
Lorentzian metric structure can emerge. In Sec.~\ref{sec:stability} we
consider a cosmological background in the effective Lorentzian theory
and analyze its stability against tensor and scalar
perturbations. Sec.~\ref{sec:discussion} is devoted to a summary 
of the results and discussions.

\section{Power-counting renormalizable theory}
\label{sec:theory}

Let us consider a $4$-dimensional Riemannian manifold ${\cal M}$ with a
positive definite metric $g^{\rm E}_{\mu\nu}$. The theory we shall consider on
this manifold does not have the concept of time. In order to make the
notion of time to emerge spontaneously, following
\cite{Mukohyama:2013ew}, let us consider a real scalar field $\phi$ with
shift symmetry. We shall see that the value of $\phi$ plays the role of
time. For this reason, we call $\phi$ a {\it clock field}. The shift
symmetry is necessary for the system after the emergence of time to have
the time translation symmetry. We also demand that the theory respects
the symmetry under the $Z_2$ transformation $\phi\to -\phi$ in order to
ensure that the system after the emergence of time has the time
reversal symmetry. We also demand that the theory is invariant under 
the $4$-dimensional parity, i.e. $x^{\mu}\to -x^{\mu}$.

We demand that the short-distance behavior of the system is dominated by
forth-order derivative terms so that the scaling dimensions of
$g^{E}_{\mu\nu}$ and $\phi$ become zero at short distances. This allows
us to construct a power-counting renormalizable theory~\footnote{In the
previous work~\cite{Mukohyama:2013ew}, to minimize the number of
physical degrees of freedom and to simplify the system, it was imposed
as a convenient assumption that equations of motion be second-order. On
the other hand, in the present paper we seek a theory which has a
potential to be renormalizable, asymptotically safe and thus UV
complete. For this reason, we consider a power-counting renormalizable
theory.} describing $g^{E}_{\mu\nu}$ and $\phi$. With the shift- and
$Z_2$-symmetries and the $4$-dimensional parity invariance mentioned
above, the system at short distances is described by the action of the
form 
\begin{eqnarray}
 I_4 & = & \int dx^4\sqrt{g_{\rm E}}
  \left[ c_1 R_{\rm E}^2 + c_2 R_{\rm E}^{\mu\nu}R^{\rm E}_{\mu\nu}
   + c_3 R_{\rm E}^{\mu\nu\rho\sigma}R^{\rm E}_{\mu\nu\rho\sigma}
   + c_4 X_{\rm E} R_{\rm E} \right. \nonumber\\
 & & 
  \left.
   + c_5 R_{\rm E}^{\mu\nu}\partial_{\mu}\phi\partial_{\nu}\phi
   + c_6 X_{\rm E}^2 + c_7 (\nabla_{\rm E}^2\phi)^2
   + c_8 (\nabla^{\rm E}_{\mu}\nabla^{\rm E}_{\nu}\phi)^2
  \right], 
\end{eqnarray} 
where $\nabla^{\rm E}_{\mu}$ , $R_{\rm E}^{\mu\nu\rho\sigma}$,  
$R_{\rm E}^{\mu\nu}$ and $R_{\rm E}$ are the covariant derivative,
Riemann curvature, Ricci curvature and Ricci scalar of the Riemannian
metric $g^{\rm E}_{\mu\nu}$,
\begin{equation}
 X_{\rm E} \equiv 
  g_{\rm E}^{\mu\nu}\partial_{\mu}\phi\partial_{\nu}\phi, \quad
  (\nabla^{\rm E}_{\mu}\nabla^{\rm E}_{\nu}\phi)^2 \equiv 
  (\nabla_{\rm E}^{\mu}\nabla_{\rm E}^{\nu}\phi)
  (\nabla^{\rm E}_{\mu}\nabla^{\rm E}_{\nu}\phi), 
\end{equation}
$g_{\rm E}^{\mu\nu}$ and $g_{\rm E}$ are the inverse and the determinant
of $g^{\rm E}_{\mu\nu}$, 
and $c_i$ ($i=1,\cdots,8$) are constants. Here, we have neglected total
derivatives. At long distances, terms with less number of derivatives
become important. There are two independent terms with two derivatives 
\begin{equation}
 I_2 = \int dx^4\sqrt{g_{\rm E}}
  \left[ c_9 R_{\rm E} + c_{10}X \right], 
\end{equation}
and a term without derivatives
\begin{equation}
 I_0 = c_{11}\int dx^4\sqrt{g_{\rm E}}, 
\end{equation}
where $c_i$ ($i=9,10,11$) are constants.

Without loss of generality, one can set $c_5=0$ by integration by parts
and redefinition of $c_{7,8}$. One can also set $c_6=1$ by rescaling
of $\phi$, provided that $c_6$ before rescaling is positive. Hence, the
total action is rewritten as 
\begin{eqnarray}
 I & = & I_4 + I_2 + I_0 \nonumber\\
 & = & 
  \int dx^4\sqrt{g_{\rm E}}
  \left[ 2Z \Lambda_{\rm E} - Z R_{\rm E} 
   + \frac{1}{2\lambda}C_{\rm E}^2 - \frac{\omega}{3\lambda}R_{\rm E}^2 
   + \frac{\theta}{\lambda}E_{\rm E}
  \right. \nonumber\\
 & & \left.
      + X_{\rm E}^2 - 2X_{\star}X_{\rm E}
      + \alpha (\nabla_{\rm E}^2\phi)^2 
      + \beta (\nabla^{\rm E}_{\mu}\nabla^{\rm E}_{\nu}\phi)^2
      + \gamma X_{\rm E} R_{\rm E} \right], \label{eqn:action}
\end{eqnarray} 
where $C_{\rm E}^2\equiv R_{\rm E}^{\mu\nu\rho\sigma}R^{\rm E}_{\mu\nu\rho\sigma}-2R_{\rm E}^{\mu\nu}R^{\rm E}_{\mu\nu}+R_{\rm E}^2/3$ is the square of the Weyl tensor, 
$E_{\rm E}\equiv R_{\rm E}^{\mu\nu\rho\sigma}R^{\rm E}_{\mu\nu\rho\sigma}-4R_{\rm E}^{\mu\nu}R^{\rm E}_{\mu\nu}+R_{\rm E}^2$
is the integrand of the Euler (or Gauss-Bonnet) term, and ($Z$,
$\Lambda_{\rm E}$, $\lambda$, $\omega$, $\theta$, $X_{\star}$,
$\alpha$, $\beta$, $\gamma$) are constants~\footnote{For the
coefficients of purely geometrical terms in the action, we adopted the
notation often used in the literature of the asymptotic safety
scenario~\cite{review-AS}. In particular, we keep the Euler term since
it ceases to be topological when some regularization scheme such as the
dimensional regularization is employed. On the other hand, the constant
$X_{\star}$ was introduced so that the polynomial of $X_{\rm E}$ in the
action is of the form $(X_{\rm E}-X_{\star})^2+const.$ as in ghost 
condensate~\cite{ArkaniHamed:2003uy,ArkaniHamed:2005gu}.
\label{footnote:notation}}.

The action (\ref{eqn:action}) is power-counting renormalizable. It
should be noted that, unlike Ho\v{r}ava-Lifshitz
gravity~\cite{Horava:2009uw}, the anisotropic scaling is not invoked
here. This is the reason why we need only up to fourth (not sixth) order
derivatives in the action. Also, the theory enjoys the four-dimensional
Riemannian diffeomorphism invariance and this significantly reduces the
number of possible terms in the action. Another difference from
Ho\v{r}ava-Lifshitz gravity is the absence of the so-called scalar
graviton, again due to the four-dimensional Riemannian diffeomorphism
invariance. Instead, in the present theory the emergence of time and
dynamics requires inclusion of the clock field $\phi$.

It is worthwhile to stress here again that the signature of the metric
is positive definite and that there is no notion of time and dynamics at 
the fundamental level. Hence, higher derivative terms do not necessarily
lead to a problem as far as the action is bounded from below.

\section{IR action}
\label{sec:IRaction}

Let us suppose a situation in which scalar invariants made of the curvature of 
$g^{\rm E}_{\mu\nu}$ and derivatives of the curvature are low in the
unit of $Z$ while $X_{\rm E}$ may become relatively large. In this
situation we can ignore the higher curvature terms $C_{\rm E}^2$,
$R_{\rm E}^2$ and $E_{\rm E}$. The system is then described by the
action (\ref{eqn:action}) without these higher curvature terms. If
$\alpha=-\beta=-2\gamma$ then this is a special case of the shift- and 
$Z_2$-symmetric, Riemannian version of the covariant Galileon action
considered in \cite{Mukohyama:2013ew} and the corresponding equations of
motion are second order. For general values of $\alpha$ and $\beta$,
however, the equations of motion include terms containing higher
derivatives. These higher derivative terms in the equations of motion
are proportional to either $\alpha+2\gamma$ or $\beta-2\gamma$.

We are allowing for a relatively large $X_E$. On the other hand, let us
suppose that scalar invariants made of second or higher covariant
derivatives of $\phi$ are small in the unit of $Z$. In this case, while
we should keep terms that stem from $\gamma X_{\rm E}R_{\rm E}$, we can
safely neglect higher derivative terms in the equations of motion, which
are proportional to either $\alpha+2\gamma$ or $\beta-2\gamma$. Hence,
the long-distance behavior of the system can be described by the
following IR action 
\begin{eqnarray}
 I_{\rm IR} & = & 
  \int dx^4\sqrt{g_{\rm E}}
  \left[ 2Z \Lambda_{\rm E} - Z R_{\rm E} \right. \nonumber\\
 & & \left.
      + X_{\rm E}^2 - 2X_{\star}X_{\rm E}
      - 2\gamma (\nabla_{\rm E}^2\phi)^2 
      + 2\gamma (\nabla^{\rm E}_{\mu}\nabla^{\rm E}_{\nu}\phi)^2
      + \gamma X_{\rm E} R_{\rm E} \right]. \label{eqn:IRaction}
\end{eqnarray} 
As already stated above, this is a special case of the shift- and
$Z_2$-symmetric, Riemannian version of the covariant Galileon action
considered in \cite{Mukohyama:2013ew} and the corresponding equations of
motion are second order~\footnote{The original Galileon theory was
proposed in ref.~\cite{Nicolis:2008in}. The covariantized version of
Galileon was then found in ref.~\cite{Deffayet:2011gz} and is equivalent
to Horndeski theory~\cite{Horndeski:1974wa}.}. (See (\ref{eqn:G4-K})
below for the precise correspondence between the IR limit
(\ref{eqn:IRaction}) of the power-counting renormalizable action and the
Riemannian Galileon action considered in \cite{Mukohyama:2013ew}.)

\section{Effective Lorentzian action}
\label{sec:LorentzianAction}

The IR action (\ref{eqn:IRaction}) (as well as the full action
(\ref{eqn:action})) is defined in terms of a Riemannian metric with
positive definite signature and thus does not have the notion 
of time at the fundamental level. However, in a region where derivative
of the clock field $\partial_{\mu}\phi$ does not vanish, one can
consider the induced metric on each constant $\phi$ hypersurface and a
sequence of the induced metrics parameterized by $\phi$. A priori, we do
not know whether the differential equation describing such a sequence is
elliptic or hyperbolic. In this and the next sections we show that the
differential equation describing the sequence of induced metrics on
constant $\phi$ surfaces can become hyperbolic when $\partial_{\mu}\phi$
is large enough. We call this phenomenon {\it emergence of time and
dynamics}.

Let us suppose that $X_{\rm E}\ne 0$ in a region ${\cal M}_0$ of the 
Riemannian manifold ${\cal M}$ with the positive definite metric 
$g^{\rm E}_{\mu\nu}$, where 
$X_{\rm E}=g_{\rm E}^{\mu\nu}\partial_{\mu}\phi\partial_{\nu}\phi$.
In this case there is a positive number $X_c$ such that $X_{\rm E}>X_c$
in ${\cal M}_0$. Under this assumption, one can define a Lorentzian
metric $g_{\mu\nu}$ with the signature ($-,+,+,+$) given by 
\begin{equation}
 g_{\mu\nu} = g^{\rm E}_{\mu\nu} 
  - \frac{\partial_{\mu}\phi\partial_{\nu}\phi}{X_c}, \quad
 g^{\mu\nu} = g_{\rm E}^{\mu\nu} 
  - \frac{g_{\rm E}^{\mu\rho}g_{\rm E}^{\nu\sigma}
  \partial_{\rho}\phi\partial_{\sigma}\phi}{X_{\rm E}-X_c}. 
  \label{eqn:def-gmunu}
\end{equation} 
As a result, we have the relation 
\begin{equation}
 \frac{1}{X} = \frac{1}{X_c}- \frac{1}{X_{\rm E}},
\end{equation}
and the inequality
\begin{equation}
 X > X_c, 
\end{equation}
where $X\equiv -g^{\mu\nu}\partial_{\mu}\phi\partial_{\nu}\phi$. 
Since we are interested in the differential equation describing the
induced metric on constant $\phi$ surfaces, changing the basic variable
from $g^{\rm E}_{\mu\nu}$ to $g_{\mu\nu}$ (defined by
(\ref{eqn:def-gmunu})) is not physically essential since these two
metrics share the same induced metric on each constant $\phi$
hypersurface. What really matters is whether the differential equation
is elliptic or hyperbolic. Nonetheless, it is still convenient for
computational purposes (e.g. for the analysis in the next section) and
also assuring to have an IR description that is manifestly covariant
with respect to the four-dimensional Lorentzian diffeomorphism. It is in
this sense that the introduction of the Lorentzian metric $g_{\mu\nu}$
is useful.

Like any gauge symmetries, (either Riemannian or Lorentzian)
diffeomorphism invariance is nothing but redundancy of description and
thus can be removed by gauge-fixing and can be restored by introduction 
of extra degrees of freedom. This simple fact suggests that there may
exist a Lorentzian description for the sequence of the induced metrics
on constant $\phi$ surfaces. Technically speaking, the construction of
such a Lorentzian description is achieved by adopting the so called
unitary gauge, i.e. by choosing one of four coordinates $t$ as 
\begin{equation}
 t = \frac{\phi}{M^2}, \label{eqn:unitarygauge}
\end{equation}
where $M$ is an arbitrary mass scale, and then undoing it. When adopting
the unitary gauge, we lose a part of the original Riemannian
diffeomorphism invariance. In the language of the Riemannian version of
the Arnowitt-Deser-Misner (ADM) formalism, it is the lapse function that
is removed (or, to be more precise, is written in terms of other 
quantities) by adopting the unitary gauge (\ref{eqn:unitarygauge}). On
the other hand, when undoing the unitary gauge, we introduce the lapse
function in the language of the Lorentzian ADM formalism. In this way we
can restore the Lorentzian diffeomorphism invariance. Hence we obtain,
so to speak, a duality between a Riemannian theory describing the
Riemannian metric $g^{\rm E}_{\mu\nu}$ and a Lorentzian theory 
describing the Lorentzian metric $g_{\mu\nu}$. The detailed derivation
of the ``duality'' can be found in \cite{Mukohyama:2013ew} and the
result is that the Riemannian action of the form 
\begin{equation}
 I_{\rm IR} = \int dx^4\sqrt{g_{\rm E}}
  \left\{ G_4(X_{\rm E})R_{\rm E} + {\cal K}(X_{\rm E})
   - 2G_4'(X_{\rm E})
   \left[ (\nabla_{\rm E}^2\phi)^2
    -(\nabla^{\rm E}_{\mu}\nabla^{\rm E}_{\nu}\phi)^2\right]
  \right\}
\end{equation}
is equivalent to the Lorentzian action
\begin{equation}
 I_{\rm IR} = \int dx^4\sqrt{-g}
  \left\{ f(X)R + P(X)
   +2f'(X)
   \left[ (\nabla^2\phi)^2
    -(\nabla_{\mu}\nabla_{\nu}\phi)^2\right]
  \right\}, \label{eqn:LorentzianAction}
\end{equation}
where $R$ and $g$ are the Ricci scalar and the determinant of the
Lorentzian metric $g_{\mu\nu}$,  and the functions $f(X)$ and $P(X)$ in
the Lorentzian action are specified as
\begin{equation}
 \frac{f(X)}{\sqrt{X}} =  \frac{G_4(X_{\rm E})}{\sqrt{X_{\rm E}}}, 
  \quad
 \frac{P(X)}{\sqrt{X}} =  \frac{{\cal K}(X_{\rm E})}{\sqrt{X_{\rm E}}}.
\end{equation}

By setting
\begin{equation}
 G_4(X_{\rm E}) = \gamma X_{\rm E} - Z, \quad
  {\cal K}(X_{\rm E}) = X_{\rm E}-2X_{\star}X_{\rm E} 
  + 2Z\Lambda_{\rm E}, \label{eqn:G4-K}
\end{equation}
we obtain 
\begin{eqnarray}
 f(X) & = & \left(\frac{\gamma X_cX}{X-X_c}-Z\right)
  \sqrt{\frac{X-X_c}{X_c}}, \nonumber\\
 P(X) & = & \left[\left(\frac{X_cX}{X-X_c}\right)^2
	     -2X_{\star}\left(\frac{X_cX}{X-X_c}\right)
	     + 2Z\Lambda_{\rm E} \right] 
 \sqrt{\frac{X-X_c}{X_c}}. \label{eqn:f-P}
\end{eqnarray}
The effective Lorentzian action (\ref{eqn:LorentzianAction}) with
(\ref{eqn:f-P}) describes the long-distance behavior of the system in
the region ${\cal M}_0$.

\section{Stability (hyperbolicity)}
\label{sec:stability}

In the previous section we have shown that the long-distance behavior of
the theory (\ref{eqn:action}) is described by the effective Lorentzian
action (\ref{eqn:LorentzianAction}) with (\ref{eqn:f-P}). This itself
does not imply the emergence of time and dynamics as defined at the
beginning of the previous section. Actually, what we need to show is
that the differential equation describing the system is hyperbolic
rather than elliptic, in some region of the Riemannian manifold. With 
the effective Lorentzian description at hand, this is equivalent to the
existence of a well-defined background around which fluctuations are
stable in the usual Lorentzian sense.

For this reason, we now analyze the stability of a flat
Friedmann-Lema\^{\i}tre-Robertson-Walker (FLRW) background using the 
effective Lorentzian action (\ref{eqn:LorentzianAction}) with
(\ref{eqn:f-P}).

\subsection{Cosmological background}

We consider a flat FLRW background spacetime for which the effective
Lorentzian metric and the clock field are
\begin{equation}
 g_{\mu\nu}dx^{\mu}dx^{\nu} = -dt^2 + a(t)^2\delta_{ij}dx^idx^j, \quad
 \phi=\phi_0(t),  \label{eqn:FLRWbackground}
\end{equation}
where $a$ is the scale factor.

The equations of motion for $\phi$ and the metric are, respectively, 
\begin{equation}
\dot{J}_{\phi} + 3H J_{\phi} = 0,  \label{eqn:eom-phi0}
\end{equation}
and 
\begin{equation}
 6 (3\gamma X_{\rm E}+Z)r H^2 = 
  3X_{\rm E}^2-2X_{\star}X_{\rm E} - 2Z\Lambda_{\rm E}, 
  \label{eqn:Friedmanneq}
\end{equation}
where $H=\dot{a}/a$, 
\begin{equation}
 J_{\phi}\dot{\phi}_0 = 
  \frac{2X_{\rm E}}{\sqrt{r}}(X_{\rm E}-X_{\star}-3\gamma r H^2),
\end{equation} 
\begin{equation}
 r \equiv \frac{X_E}{X} = \frac{X_{\rm E}-X_c}{X_c}
  = \frac{X_c}{X-X_c}, \label{eqn:def-r}
\end{equation}
and it is understood that $X_{\rm E}$ is a function of $X$ as
\begin{equation}
 X_{\rm E} = \frac{X_c X}{X-X_c}. \label{eqn:XE(X)}
\end{equation}

The equation of motion (\ref{eqn:eom-phi0}) implies that the shift
charge density $J_{\phi}$ decays as $J_{\phi}\propto 1/a^3$ and
approaches zero. By setting $J_{\phi}=0$, hence, one can obtain
equations defining attractors of the system as
\begin{eqnarray}
 (X_{\rm E}-X_{\star})
  (3\gamma X_{\rm E}-\gamma X_{\star}+2Z) 
  & = &  \gamma \left(X_{\star}^2-2Z\Lambda_{\rm E}\right), 
  \nonumber\\
 3H^2 & = & \frac{X_c}{\gamma}
  \frac{X_{\rm E}-X_{\star}}{X_{\rm E}-X_c}.
  \label{eqn:attractor}
\end{eqnarray}
These equations allow two branches of solutions: the first equation is
an algebraic equation for $X_{\rm E}$ and generically allows two
solutions. The second equation then determines the value of the Hubble
expansion rate.

If the r.h.s. of the second of (\ref{eqn:attractor}) is positive then
the universe at late time exhibits accelerated expansion. Moreover,
deviation of $J_{\phi}$ from zero generically introduces an
$O(J_{\phi})$ correction to $H^2$ and decays as 
$\propto 1/a^3$. Intriguingly, this behavior is exactly like what we
expect for dark matter. Therefore the late time behavior of the system 
is similar to that in the standard $\Lambda$CDM cosmology, at least at
the background level.

\subsection{Tensor perturbations}

We now consider tensor perturbations around the FLRW background so that
the metric and the clock field are given by 
\begin{equation}
 g_{\mu\nu}dx^{\mu}dx^{\nu} = 
  -dt^2 + a(t)^2 \left[\hbox{e}^h\right]_{ij}dx^idx^j, \quad
 \phi=\phi_0(t),  \label{eqn:tensor-perturbations}
\end{equation} 
where $h_{ij}$ is transverse and traceless
(i.e. $\partial_{i}h_{k}^i=0=\delta^{ij}h_{ij}$).

In Fourier space, the quadratic action for each polarization of the
tensor mode is given by 
\begin{equation}
 I^{(2)}_{{\rm T}, \vec{k}} = \frac{1}{8}\int dt a^3 
  \left[M_{\rm eff}^2\dot{h}_{\vec{k}}^2 - 
   2f \frac{{\vec{k}}^2}{a^2}h_{\vec{k}}^2 \right],  
\end{equation}
where
\begin{equation}
 M_{\rm eff}^2 = 2\sqrt{r}(\gamma X_{\rm E}+Z),
  \quad
  f = \frac{1}{\sqrt{r}}(\gamma X_{\rm E}-Z),  
\end{equation} 
where $r$ is defined by (\ref{eqn:def-r}) and it is again understood
that $X_{\rm E}$ is a function of $X$ as (\ref{eqn:XE(X)}). Hence, the
stability of the tensor sector requires that 
\begin{equation}
 \gamma X_{\rm E} > |Z|. \label{eqn:stability-tensor}
\end{equation}
This in particular requires that $\gamma>0$. Note that the stability
condition (\ref{eqn:stability-tensor}) applies to the system both on and
away from the attractors (\ref{eqn:attractor}).

\subsection{Scalar perturbations}

For scalar perturbations around the FLRW background, the metric and the
clock field in the unitary gauge are given by 
\begin{equation}
 g_{\mu\nu}dx^{\mu\nu} = 
  -(1 + \alpha)^2dt^2 + 2 \partial_i \beta dt dx^i
  + a(t)^2 \hbox{e}^{2\zeta}\delta_{ij} dx^idx^j, 
  \quad
  \phi = \phi_0(t). 
\end{equation}

It is straightforward to calculate the quadratic action for
perturbations, following the treatment in
ref.~\cite{Kobayashi:2011nu}. Since the time derivatives of $\alpha$ and
$\beta$ do not appear in the action, the equations of motion for
$\alpha$ and $\beta$ are constraint equations. After solving those
constraint equations with respect to $\alpha$ and $\beta$, one obtains 
the quadratic action for $\zeta$ in Fourier space as 
\begin{equation} 
 I^{(2)}_{{\rm S}, {\vec{k}}} = \frac{1}{2}\int dt a^3 
  \left[{\cal A}\dot{\zeta}_{\vec{k}}^2
   - {\cal B}\frac{{\vec{k}}^2}{a^2}\zeta_{\vec{k}}^2 \right],
  \label{eqn:quadratic-action-scalar}
\end{equation}
where ${\cal A}$ and ${\cal B}$ are given by
\begin{equation}
 {\cal A} = \frac{M_{\rm eff}^2}{H^2{\cal G}^2}
  \left(6+M_{\rm eff}^2{\cal F}\right), \quad
 {\cal B} = \frac{1}{a}
  \frac{d}{dt}\left(\frac{aM_{\rm eff}^4}{H{\cal G}^2}\right)+ 4f,
\end{equation}
${\cal F}$ and ${\cal G}$ are given by
\begin{eqnarray}
 {\cal F} & = & 
  \left[ 3X_{\rm E}^2-X_{\star}X_{\rm E}
   -3(6\gamma X_{\rm E}+Z)r H^2\right] r^{3/2},
  \nonumber\\
 {\cal G} & = & (3\gamma X_{\rm E}+Z)r^{3/2},
\end{eqnarray}
$r$ is defined by (\ref{eqn:def-r}), and it is again understood that
$X_{\rm E}$ is a function of $X$ as (\ref{eqn:XE(X)}). Hence, the
stability condition for the scalar perturbations is
\begin{equation}
 {\cal A} > 0, \quad {\cal B} > 0. \label{eqn:stability-scalar}
\end{equation}
Note that this stability condition applies to the system both on and
away from the attractors (\ref{eqn:attractor}).

Let us now study the stability condition (\ref{eqn:stability-scalar}) on
the attractors (\ref{eqn:attractor}) describing the late time
accelerated expansion of the universe. On the attractors, $X_{\rm E}$
and $H$ are constant. This means that $f$, $M_{\rm eff}^2$, ${\cal G}$
and ${\cal F}$ are also constant and that 
\begin{equation}
 {\cal B} = \frac{M_{\rm eff}^4}{{\cal G}^2} + 4f. 
\end{equation}
Hence, under the stability condition (\ref{eqn:stability-tensor}) for
the tensor sector, it follows that ${\cal B}>0$. Let us now investigate
the other stability condition ${\cal A}>0$ on the attractors. For
simplicity let us suppose that $X_{\star}=O(Z)$, that $\gamma=O(1)$,
that the expansion rate of the universe is low in the unit of $Z$
i.e. $H^2\ll |Z|$ and that $r=O(1)$. In this case, the second of the
attractor equations (\ref{eqn:attractor}) implies that $|X_{\rm
E}-X_{\star}|=O(H^2)\ll |Z|$. Hence, it is shown that 
\begin{eqnarray}
 {\cal F} & = & 
  \left[ 2X_{\star}^2 
   - \frac{1}{\gamma}\left(\gamma X_{\star}+Z\right)
   (X_{\rm E}-X_{\star})
   -3(X_{\rm E}-X_{\star})^2\right] r^{3/2} \nonumber\\
 & = & 2X_{\star}^2 r^{3/2} \left[ 1 + O(H^2/Z) \right] > 0.
\end{eqnarray}
Thus, under the stability condition (\ref{eqn:stability-tensor}) for the
tensor sector, it follows that ${\cal A}>0$.

\subsection{Summary of stability condition}

We have seen that the flat FLRW background behavior of the system is
similar to that of the standard $\Lambda$CDM cosmology. We have then
shown that tensor perturbations are stable (in the usual Lorentzian
sense), provided that the parameter $\gamma$ (defined in
(\ref{eqn:action}) or (\ref{eqn:IRaction})) is positive and that the
background value of $X_{\rm E}$ is large enough. To be more precise, the
stability condition for tensor perturbations is given by
(\ref{eqn:stability-tensor}). The stability condition for the scalar
sector shown in (\ref{eqn:stability-scalar}) is more involved but we 
have shown that it is always satisfied at the low-energy, late-time
attractor, provided that the tensor sector is stable.

The stability (in the usual Lorentzian sense) in particular implies that
the differential equation describing fluctuations of the system is
hyperbolic rather than elliptic. This is achieved by a large enough
background value of derivative of the clock filed, in the context of a
purely Riemannian theory. As defined at the beginning of the previous
section, we call this phenomenon {\it emergence of time and dynamics}.

\section{Discussion}
\label{sec:discussion}

When we talk about the history or dynamics of the universe, we are
actually taking about a sequence of configurations parameterized by
time. We often ask fundamental questions such as those concerning the
beginning of the geometrical description of the universe, and in this
case we are forced to think about the initial singularity. We might 
speculate that the notion of space does not exist before the initial
singularity and that the space may be emergent. If the space may be
emergent, then how about the time? Can the notion of time be emergent?

In any diffeomorphism invariant theories of gravity, the Hamiltonian of
the system is a sum of constraints associated with general coordinate 
transformations and thus vanishes up to boundary terms. For this reason,
there is no time evolution of quantum states in diffeomorphism invariant
theories of quantum gravity. Therefore, dynamics should be encoded as
correlations among various fields. In this case one of those fields
should play the role of time. It is perhaps in this sense that the
concepts of time and dynamics may be emergent.

In the present paper, based on the mechanism developed in
ref.~\cite{Mukohyama:2013ew}, we have proposed a new scenario of
gravitation in which gravity at short-distances is described by a
power-counting renormalizable Riemannian (i.e. locally Euclidean) theory 
without the fundamental notion of time. The Lorentzian metric structure
and the notion of time emerge as effective properties at long distances.

At the fundamental level the theory includes a Riemannian (i.e. locally
Euclidean) metric $g^{\rm E}_{\mu\nu}$ and a clock field $\phi$ playing 
the role of time. The symmetries defining the theory are the
$4$-dimensional Riemannian diffeomorphism invariance, the
$4$-dimensional parity invariance, and the shift- and $Z_2$-symmetries
of the clock field. We have written down the most general action that
contains up to forth-order derivatives. The action is shown in
(\ref{eqn:action}) and contains $9$ parameters, which are subject to
running under the renormalization group (RG) flow. Since the ultraviolet
(UV) behavior of the system is dominated by forth-order derivative
terms, the scaling dimensions of $g^{\rm E}_{\mu\nu}$ and $\phi$ are
zero in the UV and, as a result, the theory described by the action
(\ref{eqn:action}) is power-counting renormalizable.

We have then considered the infrared (IR) limit of the system and
obtained the IR action shown in (\ref{eqn:IRaction}), which turned out
to be a special case of the shift- and $Z_2$-symmetric, Riemannian
version of the covariant Galileon action. In
ref.~\cite{Mukohyama:2013ew} this IR theory was shown to be equivalent
to a Lorentzian theory in a region where the first derivative of the
clock field is non-vanishing. The theory thus has an effective
Lorentzian description valid in the IR in some regions. Therefore, the
notion of time can emerge as an effective property at long distances. On
the other hand, at short distances, forth-order derivative terms
compatible with the Riemannian diffeomorphism become important and thus 
the system is described by the power-counting renormalizable Riemannian
theory.

There are many issues to be addressed regarding theoretical consistency
and phenomenological viability of the theory.

Quantum nature of the forth-order derivative theory of gravity without
the clock field has been extensively studied in the literature. In
particular, it has been reported that the theory is
renormalizable~\cite{Stelle:1976gc}, that the dimensionless couplings
are asymptotically free~\cite{Avramidi:1985ki} and that Newton's
constant and the cosmological constant appear to be asymptotically
safe~\cite{Codello:2006in}. However, as already mentioned in 
Sec.~\ref{sec:introduction}, the Lorentzian forth-order derivative
theory has the serious issue of non-unitarity because of higher time
derivatives in the action; the challenge in the Lorentzian theory is
then how to reconcile asymptotically safe couplings with unitarity.

On the contrary, in the scenario proposed in the present paper the
theory is Riemannian (i.e. locally Euclidean) at the fundamental level
and the metric has the positive definite signature. The Lorentz metric
signature emerges as an effective property at long distances in some
regions. Thus, at short distances the notion of time and dynamics do not
exist. For this reason, the existence of higher-derivative terms is not
necessarily a problem. However, emergence of the notion of time at long
distances requires the introduction of a clock field and, as a result,
the action (\ref{eqn:action}) contains extra terms. It is thus necessary
to revisit the issues of renormalizability and asymptotic safety of the
forth-order derivative theory of gravity, with the clock field now
included.

Having a new scenario of gravitation, it is important to investigate its
cosmological implications. In the case of Ho\v{r}ava-Lifshitz
gravity~\cite{Horava:2009uw}, almost scale-invariant cosmological
perturbations can be generated even without
inflation~\cite{Mukohyama:2009gg,Mukohyama:2010xz}. The mechanism relies
on the fact that the scaling dimension of fields becomes zero in the UV,
and this property is shared by the power-counting renormalizable
Riemannian theory in the present paper, despite the fact that these two
theories are quite different. Hence, one might hope to find a similar
mechanism for generation of cosmological perturbations.

It is also intriguing to see if one can find a regime of parameters in
which the clock field behaves like ghost
condensate~\cite{ArkaniHamed:2003uy,ArkaniHamed:2005gu} and can drive
ghost inflation~\cite{ArkaniHamed:2003uz}. (See footnote
\ref{footnote:notation}.) If this is possible then the power-counting
renormalizable theory proposed in the present paper might be considered
as a possible UV completion.

In a Riemannian theory, in principle everything is determined by a
boundary condition. This may change our view on the cosmological
constant problem. In some sense, the cosmological constant problem is a
tension between the initial condition and the late time behavior of the
universe. In Lorentzian theories, quantum gravity may tell us something
about the initial condition of the universe but it is hard to imagine
how it can address the late time behavior of the universe. On the
contrary, if the notion of time is an emergent phenomenon in a
Riemannian theory then there might be a possibility that what we call 
the past and what we call the future may be ultimately related to each
other by the boundary condition that determines the whole system
including many islands with the Lorentzian metric structure as well as
many other vast regions without notion of time. However, before
addressing the cosmological constant problem, we need to develop the
quantum theory.

It is also interesting to combine the emergent time scenario with ideas
of dimensional reduction such as Kaluza-Klein compactification and
brane-world. We might end up with a landscape with various signatures
and dimensions.

While we have shown that the Lorentzian metric structure can emerge as
an effective property of the gravity sector at long distances, it is
phenomenologically important to ensure that not only the Lorentzian
signature but also the Lorentz symmetry can emerge in the matter sector
at long distances~\cite{review-LV}. For the emergence of Lorentzian
signature in the matter sector, we thus need to couple the matter sector
to derivatives of the clock field as in \cite{Mukohyama:2013ew}. It is
also necessary to develop mechanisms or symmetries to suppress Lorentz
violating operators in the matter sector at low energies after emergence
of time. It has been known that the RG flow allows a Lorentz invariant
IR fixed point so that the Lorentz symmetry emerges as a low energy
effective property of Lorentz violating
theories~\cite{Chadha:1982qq}. Although the RG running towards the
Lorentz invariant IR fixed point is typically logarithmic, it may be
possible to enhance the RG running to a power-law type by strong
dynamics~\cite{PujolasTalk}. Another possibility would be to forbid
lower dimensional Lorentz violating operators by invoking
supersymmetry~\cite{GrootNibbelink:2004za}.

\section*{Acknowledgements}
The author thanks participants of Kavli IPMU focus week on Gravity and
Lorentz violations (February 2013) for useful discussions. He is also
grateful to Jean-Philippe Uzan and Tsutomu Yanagida for helpful comments
and encouragement. This work was supported by Grant-in-Aid for
Scientific Research 24540256 and 21111006, by Japan-Russia Research
Cooperative Program and by the WPI Initiative, MEXT, Japan. 


\end{document}